\documentclass[pre,twocolumn,preprintnumbers,amsmath,amssymb,superscriptaddress]{revtex4}
\usepackage{textcomp}
\usepackage[draft]{graphicx}
\usepackage{bm}
\allowdisplaybreaks
\usepackage[breaklinks]{hyperref}
\hypersetup{colorlinks=true, linkcolor=blue, citecolor=blue, filecolor=blue, urlcolor=blue}
\usepackage{lipsum}
\usepackage[dvipsnames]{xcolor}
\usepackage{soul}
\usepackage{amsthm}
\usepackage{braket}

\newcommand{\new}{\textcolor{black}}
\newcommand{\im}{{\rm Im}}
\newcommand{\omegas}{\omega_{\rm S}}
\usepackage{enumitem}

\begin{document}

\title{Thermodynamic performance of a periodically driven harmonic oscillator correlated with the baths}

\author{Tianqi Chen}
\affiliation{Science, Mathematics and Technology Cluster, Singapore University of Technology and Design, 8 Somapah Road, 487372 Singapore}

\author{Dario Poletti}
\affiliation{Science, Mathematics and Technology Cluster, Singapore University of Technology and Design, 8 Somapah Road, 487372 Singapore}
\affiliation{Engineering Product Development Pillar, Singapore University of Technology and Design, 8 Somapah Road, 487372 Singapore}

\begin{abstract}
We consider a harmonic oscillator under periodic driving and coupled to two harmonic-oscillator heat baths at different temperatures. We use the thermofield transformation with chain mapping for this setup, which allows us to study the unitary evolution of the system and the baths up to \new{a time when the periodic steady state emerges in the system.}
We characterize this periodic steady state, and we show that, by tuning the system and the bath parameters, one can turn this system from an engine to an accelerator or even to a heater. The possibility to study the unitary evolution of the system and baths also allows us to evaluate the steady correlations that build between the system and the baths, and correlations that grow between the baths.
\end{abstract}

\maketitle

\section{Introduction}

\begin{figure}[ht]
\centering
\includegraphics[width=1.0\columnwidth, draft=false]{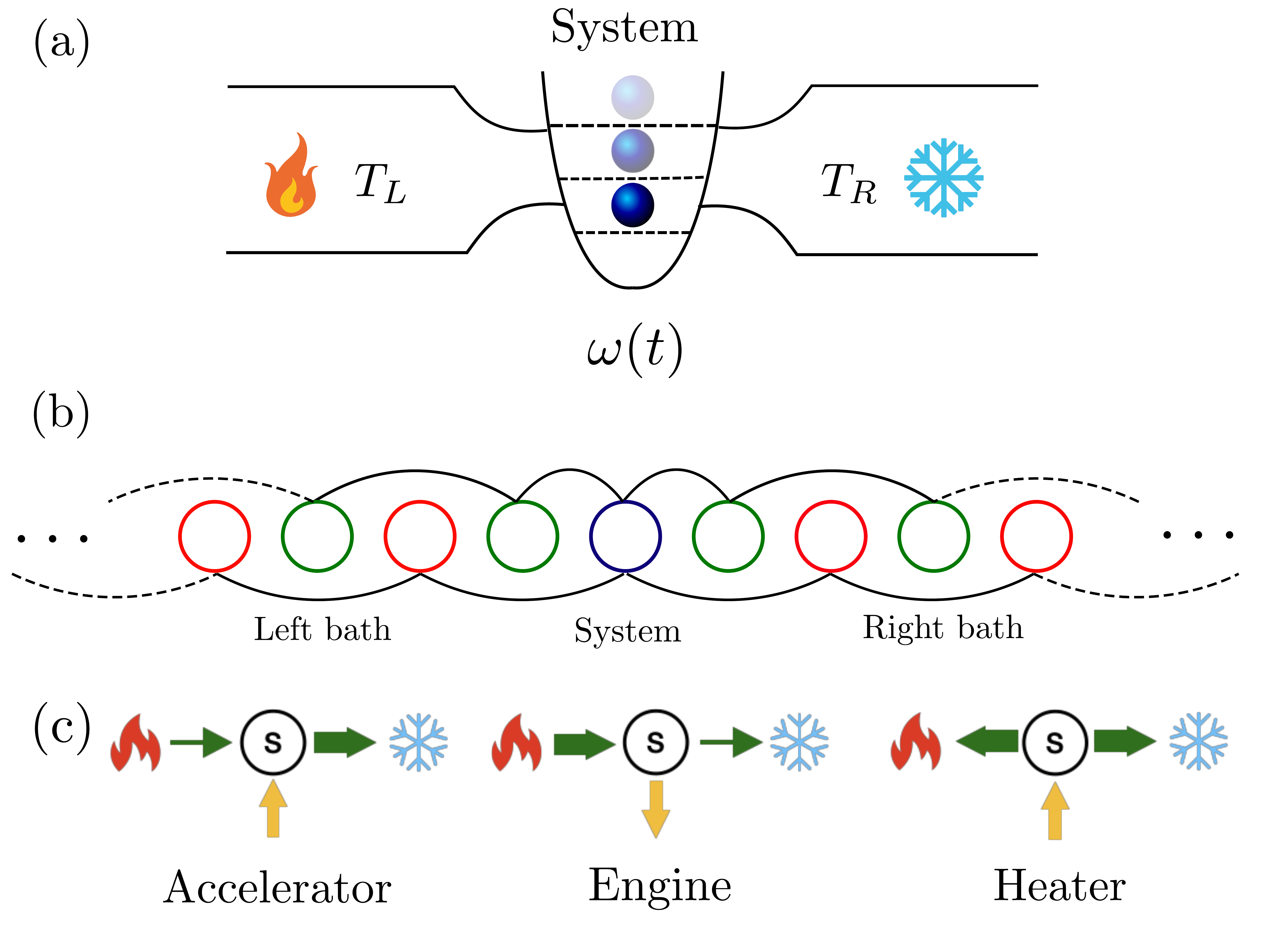}
\caption{\label{fig:fig1totalsystem} (a) Schematics of a driven single bosonic harmonic oscillator coupled to two heat baths at different temperatures. The system has a periodically varying trapping potential $\omega(t)$. The hot bath is on the left-hand side at temperature $T_L$, and the cold bath is on the right-hand side at temperature $T_R$. (b) System coupled to the two baths after thermofield transformation with chain mapping: each bath is mapped to a chain with next-nearest neighbor tunneling or, equivalently, to two chains. (c) Illustration of energy transfer patterns in accelerator, engine, and heater. }
\end{figure}

A better understanding of transport and thermodynamics at the nanoscale can potentially lead to significant technological advances \cite{Ventra2011, Dhar2008, Whitney2017, LandiSchaller2021, Goold2016, Kosloff2013, MillenXuereb2016, BroeckEsposito2015, VinjanampathAnders2016, GelbwaserJurizki2015, DeffnerCampbell2019}. In particular, the study of prototypical quantum systems can allow a very accurate modeling that uses as few approximations as possible, thus leading to deeper insights. Given a setup consisting of a system coupled to two baths at different temperatures, which results in an energy current, one can wonder whether work can be extracted from it. One way to do so is to periodically drive the system as shown in Fig.~\ref{fig:fig1totalsystem}(a), where a harmonic oscillator with a time-dependent trapping potential is coupled to the left to a hot bath with temperature $T_L$ and to the right to a cold bath at temperature $T_R$.

The modeling of these periodically driven systems coupled to thermal bath(s) is an active research topic. For weak enough system-bath coupling, for baths with fast enough decaying correlation, and for an uncorrelated system and baths at the initial time, one can model these systems with master equations in Gorini-Kossakowski-Sudarshan-Lindblad form \cite{GoriniSudarshan1976, Lindblad1976, OQSBook}. Within this approach, one only needs to deal with a master equation for the system alone. If the steady state exists and is unique, one can show that it will be periodic with the period of the driving \cite{HartmannHanggi2017}, and one can define heat and work transfers that result in consistent thermodynamic equations \cite{Alicki1979}.
However, this and similar approaches, e.g., based on Redfield master equations \cite{Redfield1965}, do not allow us to exactly study the correlations that build between the system and the baths, or even between the baths. Importantly, the effect of the baths may not be additive \cite{GiusteriCelardo2017, MitchinsonPlenio2018, MaguireNazir2019, CygorekGauger2021, LandiSchaller2021}.
Some of the recent works that have investigated thermodynamic devices such as heat-to-work and work-to-work converters beyond the limits of Markovian treatments are in Refs.~\cite{Gelbwaser2015, Brenes2020, Wiedmann2020, BenentiCangemi2020, BenentiCangemi2021}.

One approach that goes beyond the models limitations described above is to simulate both the system and the baths. While it may seem an impossible task to simulate the baths, this can be done, for example, by mapping the bath of harmonic oscillators to a semi-infinite chain (so called chain mapping) \cite{Wilson1975, Plenio2010a, ChinPlenio2010, ChinPlenio2014}. As one can only simulate finite chains, this approach gives a very accurate description of the baths up to the times at which finite-size effects become important. The chain mapping can be further assisted by a thermofield transformation which maps each bath into two zero-temperature baths \cite{deVegaBanuls2015}. Combining the thermofield transformation with chain mapping, each bath is represented by two empty semi-infinite chains \cite{deVegaBanuls2015, LandiSchaller2021}. The combination of these approaches is being used more often in the literature \cite{ChenPoletti2020, XuPoletti2019, MascarenhasVega2017, DelftArrigoni2018, DelftWeichselbaum18, de2015discretize, guo2018stable}. A depiction of the mapping is given in Fig.~\ref{fig:fig1totalsystem}(b).

It is important to stress, however, that it is not sufficient to have a periodically driven system coupled to two baths at different temperatures to obtain an engine. In effect, a thermal machine can have different thermodynamic behaviors \cite{BuffoniCampisi2019}. We illustrate three of these behaviors in Fig.~\ref{fig:fig1totalsystem}(c), in which we show different flow directions of the energy and their thermodynamics interpretation. If the energy current flows from the hot bath to the cold bath, but the current going to the cold bath is larger than the one from the hot bath because some work is done on the system, then the system is considered to be an {\it accelerator}. If instead the energy from the hot bath is larger than the one going into the cold bath, then work is done by the system and it behaves as an {\it engine}. Finally, if energy flows from the external driving to each of the baths, then the system is a {\it heater} \cite{fn1}. For accelerators and heaters, due to the direction of the energy flow between the system and the external driving, the total energy of the system and baths will increase, while for engines the total energy will decrease.

In this work, we are able to show that even in the unitary evolution of the system plus baths, \new{an approach that goes beyond weak coupling and master equations treatments,} the system goes into a periodic steady state \new{that oscillates at the frequency of the driving}. We then characterize such a periodic steady state, and we demonstrate that by tuning the system and bath parameters, one can readily turn the system from a heater, to an accelerator, and to an engine. Lastly, \new{thanks to the unitary evolution of both the system and the baths, we can witness the emergence of correlations between the system and baths, and steady growth of correlations between the baths. This highlights that thermodynamic engines can function even in regimes beyond weak coupling}.
The reminder of this article is arranged as follows: In Sec.~\ref{sec:modelsandmethods} we introduce our setup of the driven harmonic oscillator coupled to two heat baths. We discuss the spectral density of the baths used, and we present a brief review of the thermofield transformation with the chain mapping method. In Sec.~\ref{sec:results} we study the effect of both system and bath properties on the periodic steady-state dynamics of the driven system. In particular, we show the emergence of a periodic steady state, and we demonstrate that the system can act as an engine, an accelerator, or a heater. This can be altered by varying  system and bath parameters, such as the temperature, the spectral density, the amplitude of the driving, and the system-bath coupling strength. Finally, we show that correlations build up between the system and the bath, and also that growing correlations form between the baths. We draw our conclusions in Sec.~\ref{sec:conclusions}.

\section{Model and methods}
\label{sec:modelsandmethods}
\subsection{Setup and bath spectral density}
We consider a single-site bosonic harmonic oscillator coupled to two heat baths at different temperatures (Fig.~\ref{fig:fig1totalsystem}). The harmonic oscillator, which we refer to as the system in the following, has a trapping frequency $\omega(t)=\omegas + \delta\sin(\omega_{\rm dri}t)$, where $\omegas$ is a constant component, while $\delta \sin(\omega_{\rm dri}t)$ contributes to the external sinusoidal oscillation characterized by the frequency $\omega_{\rm dri}$ and amplitude $\delta$. Each bath is composed by a set of independent harmonic oscillators, and each has energy $\omega_i$. The total system plus bath Hamiltonian, $\hat{\mathcal{H}}_{\rm tot}$, can thus be expressed as the sum of the system bare Hamiltonian $\hat{\mathcal{H}}_{\rm S}$, the external driving Hamiltonian $\hat{\mathcal{H}}_{\rm EXT}$, the baths Hamiltonian $\hat{\mathcal{H}}^L_{\rm B}$, $\hat{\mathcal{H}}^R_{\rm B}$ (respectively for the left and right bath), and the system-bath interaction term $\hat{\mathcal{H}}_{\rm I}$
\begin{align}
\label{eq:totalHambeforeTC}
&\hat{\mathcal{H}}_{\rm tot} = \hat{\mathcal{H}}_{\rm S}+\hat{\mathcal{H}}_{\rm EXT}(t)+\hat{\mathcal{H}}^L_{\rm B}+\hat{\mathcal{H}}^R_{\rm B}+\hat{\mathcal{H}}_{\rm I} \\ \nonumber
&\hat{\mathcal{H}}_{\rm S} = \hbar\omega_{\rm S} a_{\rm S}^{\dagger}a_{\rm S} \\ \nonumber
&\hat{\mathcal{H}}_{\rm EXT}(t) = \hbar\delta \sin(\omega_{\rm dri} t)a_{\rm S}^{\dagger}a_{\rm S}  \\ \nonumber
&\new{\hat{\mathcal{H}}^{\nu}_{\rm B}= \sum_k \hbar\omega_k b_k^{\nu\dagger}b_k^\nu}\\ \nonumber
&\hat{\mathcal{H}}_{\rm I}= \sum_{\nu=L,R}\sum_k \hbar\omegas \sqrt{\mathcal{J}_k^\nu} \left( a_{\rm S} b_k^{\nu \dagger}+a_{\rm S} ^{\dagger}b_k^{\nu}\right).
\end{align}
Here $a_{\rm S}$ is the annihilation operator of the system, and $b_{k}^\nu$ is the annihilation operator of the bath. $\mathcal{J}^\nu_k=\int_{\omega_k}^{\omega_{k+1}}\,d\omega \mathcal{J}^\nu(\omega)\approx \mathcal{J}^\nu(\omega_k) \Delta \omega$ is the discretized spectral density of the bath $\mathcal{J}^\nu(\omega)$ \cite{fn2}.
Throughout our work, we consider a Lorentzian-type bath spectral density which can be expressed as \cite{Kofman1994}
\begin{align}
\label{eq:bathspectraldensity}
&\mathcal{J}^\nu(\omega)=\frac{\gamma_\nu}{\pi}\frac{\left[\pi G_\nu(\omega)\right]^2}{(\omega-\omega^\nu_r)^2+\left[\pi G_\nu(\omega)\right]^2}\,\,(\nu=L,R)
\end{align}
where $\gamma_\nu$ is a dimensionless system-bath coupling strength, $\omega^\nu_r$ is the resonance frequency for each bath spectrum, and $G_\nu(\omega)=A_\nu \omega$ (ohmic system-bath coupling spectrum at low frequencies), \new{and $A_{\nu}$ is a constant of which the magnitude controls the width of the Lorentzian-type bath spectral density: the larger $A_{\nu}$ is, the wider is the spectral density function}. \new{Such a spectral function for the bath can be produced, for instance, with a single mode of a cavity with a finite linewidth \cite{Kofman1994, KofmanKurizki1996}}. In the following, we will use $\gamma_L=\gamma_R=\gamma$. Effectively, this spectral density can be seen as imposing a filter onto each system-bath coupling spectrum such that the system is coupled more strongly to modes closer to the resonance frequency $\omega^\nu_r$. Finally, $\mathcal{J}^\nu(\omega)=0$ for $\omega$ larger than the cut-off frequency $\omega_c$.
\new{We point out that in the following we focus on using $\omega_{\rm dri} = 0.5 \omega_S$ as this allows us to get a steady state in about four periods of the driving, thus making the simulations more manageable. }

\subsection{Equations of motion after thermofield transformation with chain mapping} \label{sec:TCEOM}
Here we present the method we use in this work for the time evolution of the total system to obtain the periodic steady state. We apply a unitary, numerically exact, thermofield transformation with chain mapping as described in Ref.~\cite{ChenPoletti2020} to both bath Hamiltonians in Eq.~\eqref{eq:totalHambeforeTC}. Each bath is mapped onto two zero-temperature decoupled tight-binding chains with nearest-neighbor tunneling coefficients $\beta_{\zeta,j}^\nu$, and the on-site potential $\alpha_{\zeta,j}^\nu$, where $\nu=L,R$ denotes the index for the bath. $\zeta=1,2$ is the index for each tight-binding chain of the bath, and $j$ is a discrete index for the site numbers.
The final form of the total Hamiltonian after the transformations, $\mathcal{\hat{H}}^{\text{TC}}_{\text{tot}}$, then becomes
\begin{align}
\label{eq:totalHam}
&\mathcal{\hat{H}}^{\text{TC}}_{\text{tot}}/\hbar\omegas = a_{\rm S}^\dagger a_{\rm S} + \left(\delta/\omegas\right) \sin(\omega_{\rm dri}t)a_{\rm S}^\dagger a_{\rm S} \\
\nonumber &+ \sum_{\zeta,\nu} \left[  \sum_{j=1}^{N_{\text{chain}}} \!\!\! \alpha^{\nu}_{\zeta,j} d^{\nu\dagger}_{\zeta,j}d^{\nu}_{\zeta,j} + \!\!\!\!\!\!\sum_{j=1}^{N_{\text{chain}}-1} \!\!\!\!\!\! \beta^{\nu}_{\zeta,j}\left(d^{\nu\dagger}_{\zeta,j}d^{\nu}_{\zeta,j+1}+d^{\nu}_{\zeta,j}d^{\nu\dagger}_{\zeta,j+1} \right) \right] \\ \nonumber &+\sum_{\nu}\left[\beta^{\nu}_{1,0} \left(a_{\rm S}^{\dagger} d^{\nu}_{1,1} + a_{\rm S} d^{\nu\dagger}_{1,1} \right)+\beta^{\nu}_{2,0} \left(a_{\rm S}^{\dagger} d^{\nu \dagger}_{2,1} + a_{\rm S} d^{\nu}_{2,1} \right)\right].
\end{align}
\textcolor{black}{with $d_{\zeta,j}$ being the annihilation operator of the bath after the transformation.} If we align the total system in one dimension as in \cite{ChenPoletti2020}, the above Eq.~\eqref{eq:totalHam} is essentially a non-interacting chain of harmonic oscillators with next-nearest neighbor tunneling, only with non-number conserving interaction terms between the system site and the first site of the bath chain labeled by $\zeta=2$.  The length of the total chain is ${ N_{\text{tot}}}=4\times{ N_{\text{chain}}}+1$ \cite{fn_length_chain}.
\new{We evolve the equations of motion for both the system and the baths, and we examine the observables to see whether the periodic steady state emerges or not. In this paper, for each set of the system and the bath parameters used, we have evolved the total system for a certain amount of time, and we checked that between the last two periods the instantaneous difference from an observable, e.g., the system occupation, at time $t$ and $t+2\pi/\omega_{\rm dri}$ is smaller than $10^{-5}$.}

\section{Results}
\label{sec:results}

In this section, we study the periodic steady-state transport and thermodynamic properties of the system and the baths.
We will first show the emergence of a steady state in Sec.~\ref{sec:periodic_ss}, discuss the role of the spectral densities in Sec.~\ref{sec:separationofbath}, show how this results in different qualitative types of thermodynamic performance of the system in Sec.~\ref{sec:classification}, and study the correlations formed in Sec.~\ref{sec:correlations}.
In the following, the temperature is given in units of $T_0=\hbar\omegas/k_B$ where $k_B$ is the Boltzmann constant.

\subsection{Periodic steady-state system occupation and energy currents from the bath}
\label{sec:periodic_ss}

\begin{figure}[ht]
\centering
\includegraphics[width=1.\columnwidth, draft=false]{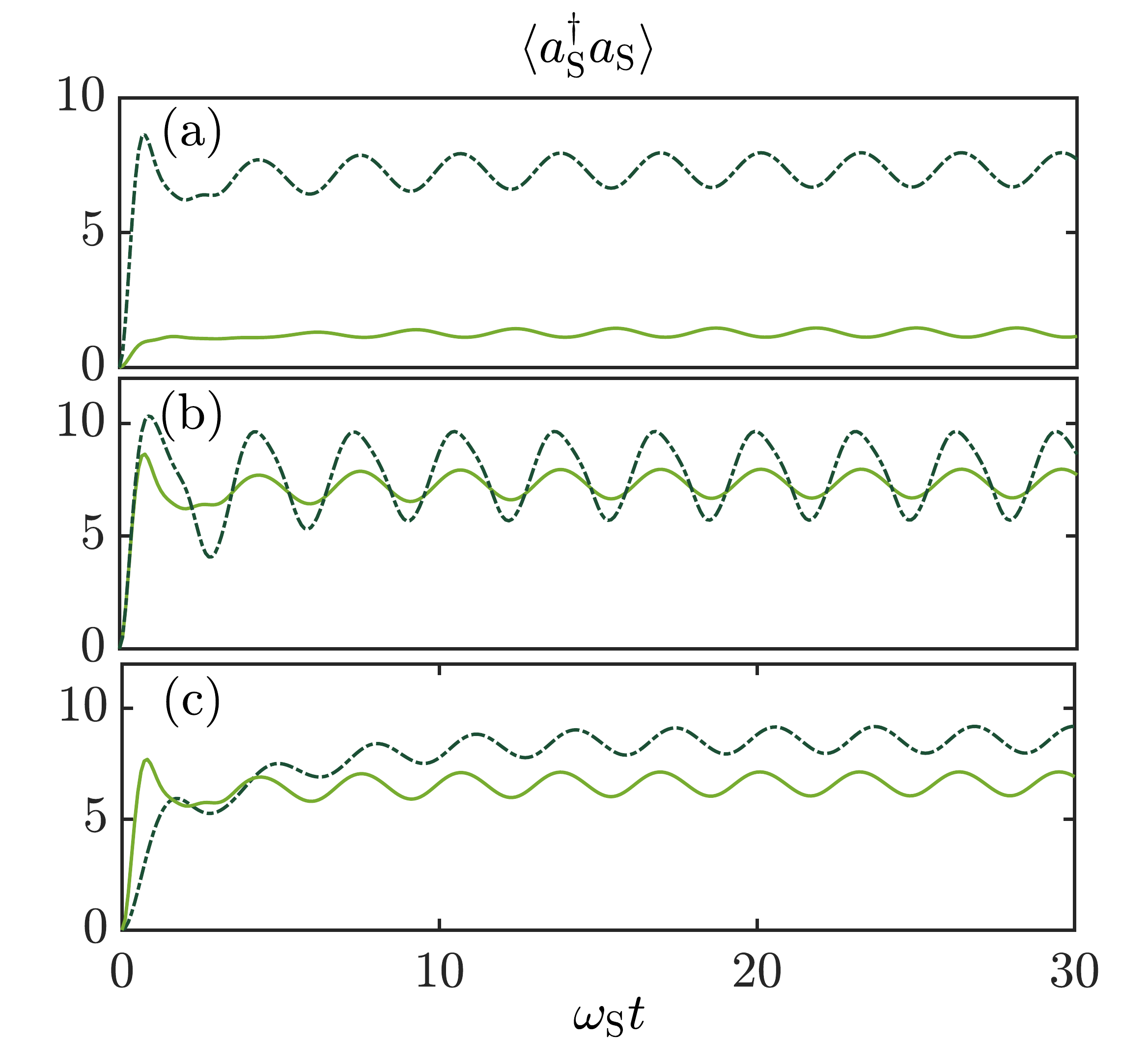}
\caption{\label{fig:occupation} \new{Periodic system occupation between $\omegas t=0$ and $\omegas t=30$ for higher and lower (a) temperatures: $T_L=2T_0$, and $T_L=22.5T_0$. Other parameters are: $\gamma=1,\delta=0.1\omegas$; (b) driving amplitude: $\delta=0.1\omegas$, and $\delta=0.5\omegas$. Other parameters are: $T_L=22.5T_0$, $\gamma=1$; and (c) system-bath coupling strength: $\gamma=0.2$, and $\gamma=1$. Other parameters are: $T_L=20T_0$, $\delta=0.1\omegas$. In all panels, the lighter solid curve is for the smaller parameter, and the darker solid-dashed curve is for the larger parameter, and we have used $T_R=T_0$, $\omega_{\rm dri}=0.5\omegas$, $\omega_c=5\omegas$, $A_L=A_R=0.0125$, $\omega^L_r=1.5\omegas$, $\omega^R_r=0.5\omegas$.}}
\end{figure}

\begin{figure}[ht]
\centering
\includegraphics[width=1.\columnwidth, draft=false]{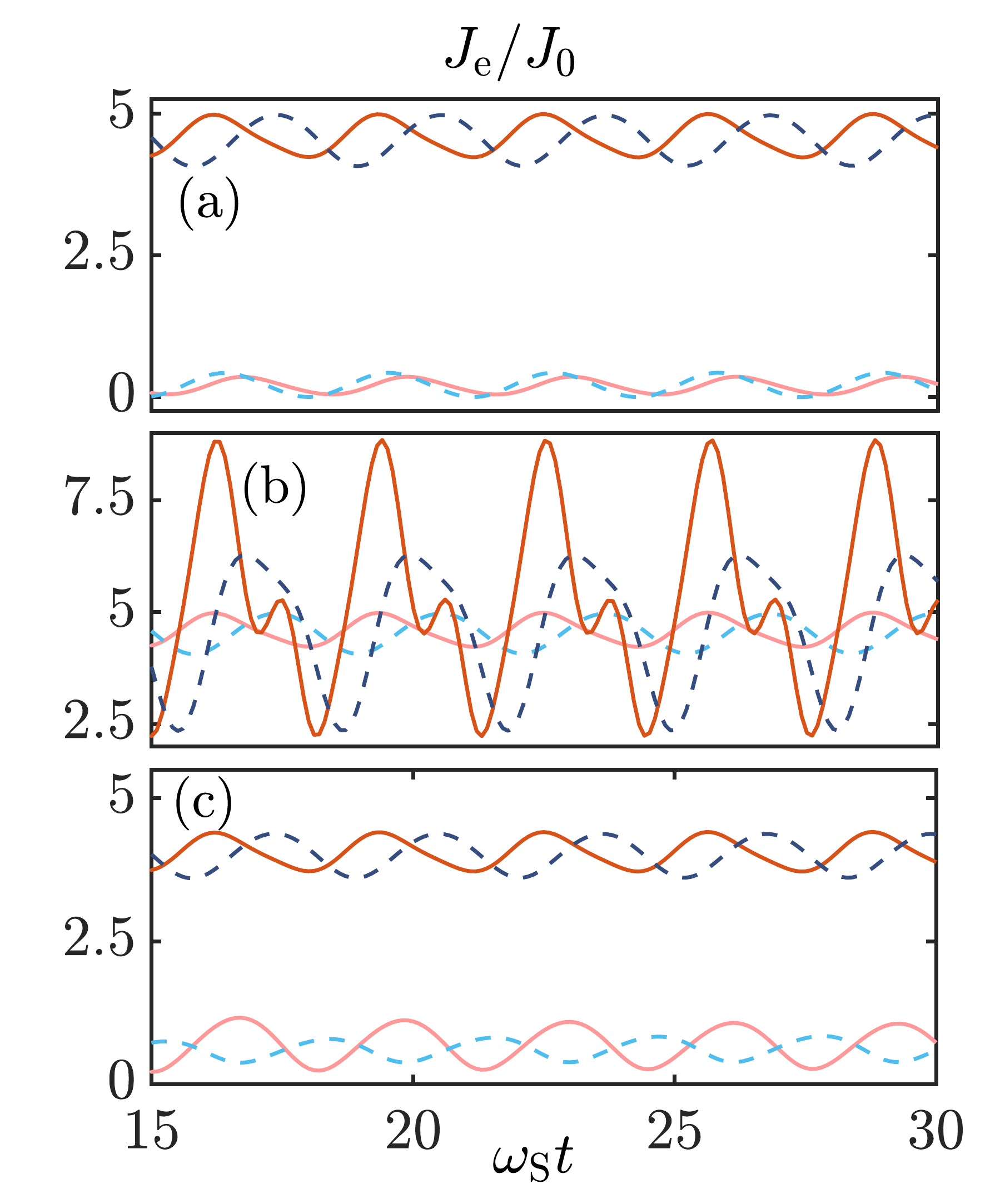}
\caption{\label{fig:current} \new{Time evolution of the energy currents from baths between $\omegas t=15$ and $\omegas t=30$. Here, the solid red lines represent the energy current from the left bath, and the dashed blue lines represent the energy current to the right bath. (a) The lighter curves are for the left bath temperature $T_L=2T_0$, and the darker curves are for the left bath temperature $T_L=22.5T_0$. Other parameters are: $\gamma=1,\delta=0.1\omegas$. (b) The lighter curves are for the driving amplitude $\delta=0.1\omegas$, and the darker curves for $\delta=0.5\omegas$. Other parameters are: $T_L=22.5T_0$, $\gamma=1$. (c) the lighter curves are for the system-bath coupling strength $\gamma=0.2$, and the darker curves are for the system-bath coupling strength $\gamma=1$. Other parameters are: $T_L=20T_0$, $\delta=0.1\omegas$. For all panels, we have used $T_R=T_0$, $\omega_{\rm dri}=0.5\omegas$, $\omega_c=5\omegas$, $A_L=A_R=0.0125$, $\omega^L_r=1.5\omegas$, $\omega^R_r=0.5\omegas$.} }
\end{figure}


We utilize the Heisenberg equation of motion to derive the energy currents from the left $L$ and right $R$  baths
\begin{align}
\label{eq:bathcurrentEOM}
&J_e^L=-\frac{d}{dt}\langle \hat{\mathcal H}_{\rm B}^L \rangle = -\frac i \hbar \left\langle \left[\hat{\mathcal{H}}_{\rm tot}^{}, \hat{\mathcal H}_{\rm B}^L\right]\right\rangle \\ \nonumber
&J_e^R=\frac{d}{dt}\langle \hat{\mathcal H}_{\rm B}^R \rangle = \frac i \hbar \left\langle \left[\hat{\mathcal{H}}_{\rm tot}^{}, \hat{\mathcal H}_{\rm B}^R\right]\right\rangle.
\end{align}
The above different signs for the left and the right bath indicate that the positive direction of the energy flow is from the left bath into the system, and from the system into the right bath. The energy currents can thus be expressed in terms of correlation functions after the thermofield transformation with chain mapping as
\begin{align}
\label{eq:bathcurrentexpression}
J_e^L/J_0 &= -\left(2\beta_{1,0}^L\beta_{1,1}^L \im\langle d_{1,2}^{L\dagger}d_{\rm S}  \rangle +2\beta_{1,0}^{L}\alpha_{1,1}^L\im\langle d_{1,1}^{L\dagger}a_{\rm S}\rangle \right. \nonumber \\ \nonumber
& \left.+2\beta_{2,0}^{L}\beta_{2,1}^L\im\langle d_{2,2}^{L\dagger}a_{\rm S} \rangle +2\beta_{2,0}^L\alpha_{2,1}^L\im \langle d_{2,1}^{L\dagger}a_{\rm S}\rangle \right)  \\ \nonumber
J_e^R/J_0 &= \left( 2\beta_{1,0}^R\beta_{1,1}^R \im\langle d_{1,2}^{R\dagger}a_{\rm S}  \rangle +2\beta_{1,0}^{R}\alpha_{1,1}^R\im\langle d_{1,1}^{R\dagger}a_{\rm S}\rangle \right. \\
& \left.+2\beta_{2,0}^{R}\beta_{2,1}^R\im\langle d_{2,2}^{R\dagger}a_{\rm S} \rangle +2\beta_{2,0}^R\alpha_{2,1}^R\im \langle d_{2,1}^{R\dagger}a_{\rm S}\rangle \right)
\end{align}
where we have expressed the energy current in the unit $J_0 = \hbar \omegas^2$.
The periodic behavior of the system occupation $\langle a_{\rm S}^\dagger a_{\rm S}\rangle$ and the energy current $J_e^L, J_e^R$ versus time is plotted in Fig.~\ref{fig:occupation} \new{and Fig.~\ref{fig:current} respectively} for different bath temperature bias, driving amplitude, and system-bath coupling strength.
For all the parameters considered, the occupation and the energy currents show a periodic steady-state behavior with a period of $2\pi/\omega_{\rm dri}$, indicating that the time evolution is sufficiently long enough. If longer times are needed to reach the steady state, then one needs to consider longer chains in the chain mapping of the baths.

In \new{Fig.~\ref{fig:occupation}(a) and Fig.~\ref{fig:current}(a)}, the $\langle a_{\rm S}^\dagger a_{\rm S}\rangle$ and $J_e^L$ and $J_e^R$ versus time are plotted for different temperatures of the hot bath on the left, $T_L$, while the cold bath is kept at $T_R=T_0$ (darker curves represent larger bias). We observe that for larger $T_L$, both the system occupation and the energy current increase. \new{This can be understood as the result of an increase of bias between the two baths.}
In \new{Fig.~\ref{fig:occupation}(b) and Fig.~\ref{fig:current}(b)}, the system occupation and the energy currents are studied for different driving amplitudes $\delta$. It is observed that the amplitudes for both the system occupation and the energy currents increase for larger $\delta$. Notice that for very strong $\delta=0.5\omegas$ here (darker curves for stronger driving), the curves of energy currents are not simple sinusoidal functions, yet they preserve the period of the driving $2\pi/\omega_{\rm dri}$. \new{This can be understood by the fact that larger amplitude oscillations result in stronger non-linear response of the system, thus a non-sinusoidal evolution in time.}
In \new{Fig.~\ref{fig:occupation}(c) and Fig.~\ref{fig:current}(c)} we show that an increase in the system-bath coupling $\gamma$ from Eq.~(\ref{eq:bathspectraldensity}) can result\new{, as it can be expected in linear response,} in an increase in the occupation of the oscillator and the current between the baths.

\subsection{Separation of bath spectral densities}
\label{sec:separationofbath}

\begin{figure}[ht]
\centering
\includegraphics[width=1.0\columnwidth, draft=false]{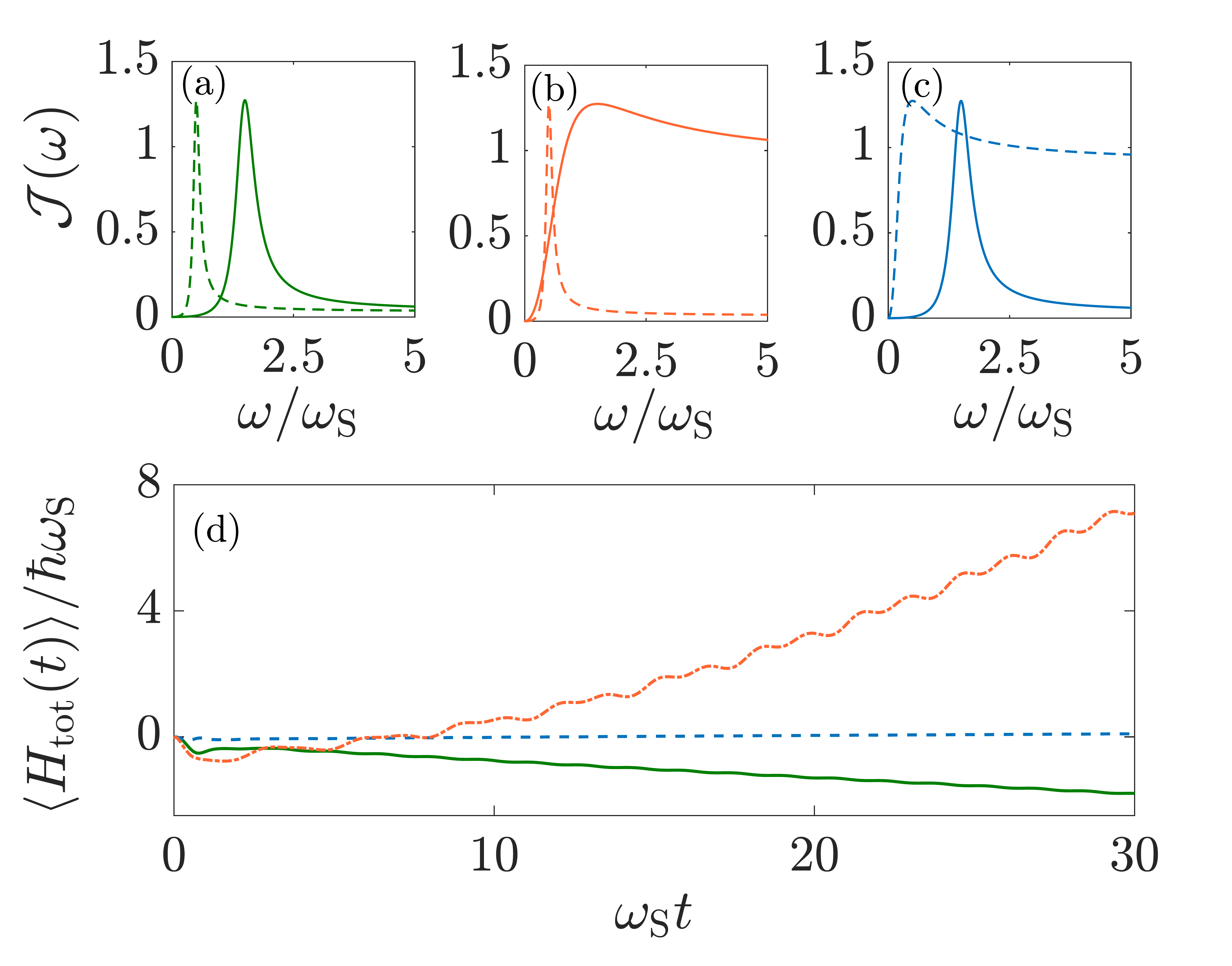}
\caption{\label{fig:spectraldensityseperation} Comparison of the effect of the separation of the bath spectral densities and the total energy: (a), (b), and (c) Lorentzian-type bath spectral density. (a)$A_L=A_R=0.0125$ (green); (b)$A_L=0.125$, $A_R=0.0125$ (orange); (c)$A_L=0.0125$, $A_R=0.125$ (blue). From panel (a) to (c), solid curves are for the left bath spectral density, and the dashed lines are for the right bath spectral density. (d) The total energy of the system and two baths as a function of time. The green solid, orange dot-dashed, and blue dashed curves correspond, respectively, to the bath spectral densities used in panels (a), (b) and (c). Other parameters used are: \textcolor{black}{$\gamma=1$,} $T_L=20T_0$, $T_R=T_0$, $\omega_c=5\omegas$, $\omega_{\rm dri}=0.5\omegas$, $\omega^L_r=1.5\omegas$, $\omega^R_r=0.5\omegas$, $\delta=0.1\omegas$.}
\end{figure}

We now proceed to focus on the thermodynamic properties of our total system. Since we can simulate the total setup of system plus baths, we can also monitor its total energy, which, since the total Hamiltonian is time-dependent, can change in time. In particular, if one observes a steady increase of the energy, it would mean that the external driving is doing work, while if the total energy is constantly decreasing, it would mean that the system is doing work, thus behaving like an engine.
In Fig.~\ref{fig:spectraldensityseperation}(d) we observe that the green continuous line reflects the emergence of engine behavior. This curve corresponds to the spectral densities of the baths given in Fig.~\ref{fig:spectraldensityseperation}(a), which are well separated.
\new{This bath spectral density separation indicates that the harmonic oscillator in the system is more strongly coupled to the hot bath only at high frequencies, while it is more strongly coupled to the cold bath only at lower frequencies. In such a case, thanks to the driving, there can be energy transfer from the hotter and higher frequency bath (which is also more highly populated), to the cooler and lower frequency one (which is less populated), but not vice versa}.
For the cases in which the spectral densities are overlapping, see Fig.~\ref{fig:spectraldensityseperation}(b) and (c), one cannot readily obtain engine behavior.
Instead, the overall energy can be almost constant or increase rapidly,
\new{as the above condition for the bath spectral density separation no longer exists, and therefore the system is now coupled to a variety of modes from each one of the baths.}
The three scenarios of spectral densities are produced by changing the magnitude of $A_\nu$ in $G_\nu(\omega)=A_\nu \omega$, which goes into Eq.~(\ref{eq:bathspectraldensity}), while we have kept the resonant frequencies $\omega^\nu_r$ fixed.
\new{We highlight that, unlike in the transport problem, it is not necessary for the spectral functions of the baths to have any overlap. This is because the driving is going to ``bridge the gap'' and couple them if they are separated in the frequency domain. }

\subsection{Tuning the system's behavior as thermal engine, heater or accelerator}
\label{sec:classification}

\begin{figure}[ht]{}
\centering
\includegraphics[width=1.0\columnwidth, draft=false]{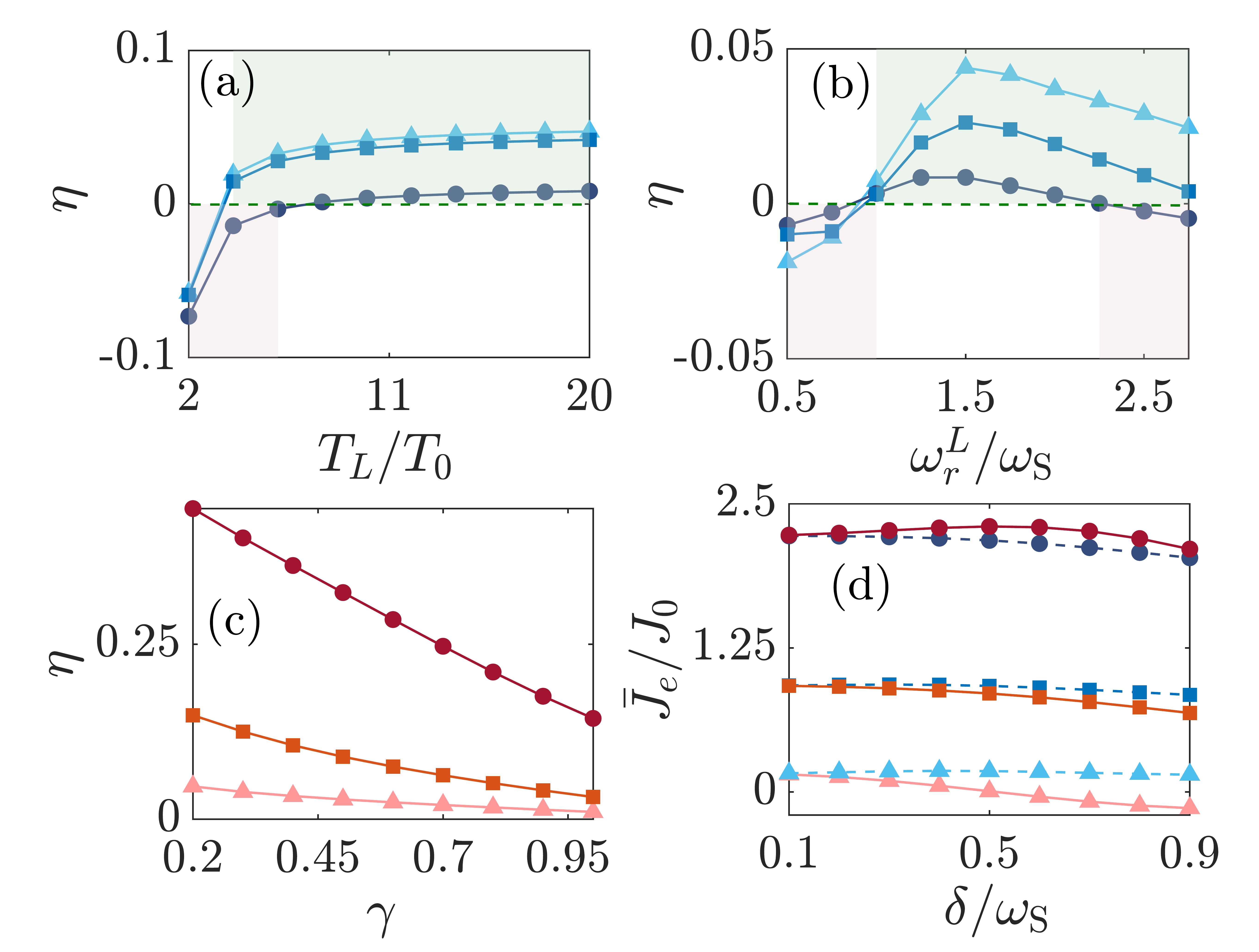}
\caption{\label{fig:classification} Thermal machine efficiency and averaged currents: (a) Efficiency $\eta$ as a function of the left bath temperature $T_L$ with ascending system-bath coupling strength $\gamma=0.2$, $0.4$, $1$ (darker color for larger value). Other parameters used are: $T_R=T_0$, $\omega_c=5\omegas$, $\omega^L_r=1.5\omegas$, $\omega^R_r=0.5\omegas$, $\delta=0.1\omegas$, and $A_L=A_R=0.0125$. (b) Efficiency $\eta$ as a function of the left bath spectral density resonant frequency $\omega^L_r$ with ascending system-bath coupling strength $\gamma=0.25$, $0.5$, $1$ (darker color for larger value). Other parameters used are: $T_L=20T_0$, $T_R=T_0$, $\omega_c=5\omegas$, \textcolor{black}{$\omega^R_r=0.5\omegas$}, $ \delta=0.1\omegas$, and $A_L=A_R=0.0125$. In (a) and (b) the shaded pink region indicates the regime for accelerator, and the shaded green region for engine. (c) Efficiency as a function of system-bath coupling strength $\gamma$ with ascending driving amplitude $\delta=0.1\omegas$, $0.2\omegas$, $0.6\omegas$ (darker color for larger value). Other parameters used are: $T_L=20T_0$, $T_R=T_0$, $\omega_c=5\omegas$, $\omega^L_r=1.5\omegas$, $\omega^R_r=0.5\omegas$, $A_L=A_R=0.0125$. (d) Averaged energy currents as a function of driving amplitude $\delta$ with ascending left bath temperature $T_L=1.5T_0$, $3.75T_0$, $7.5T_0$ (darker color for larger value). The solid lines are for the energy current from the left bath, and the dashed lines for the right bath. Other parameters used are: $T_R=T_0$, $\gamma=0.5$, $\omega_c=5\omegas$, $\omega_{\rm dri}=0.5\omegas$, $\omega^L_r=1.5\omegas$, $\omega^R_r=0.5\omegas$, $A_L=A_R=0.0125$. }
\end{figure}

As we could observe in Fig.~\ref{fig:spectraldensityseperation}(d), the system may behave as an engine, when the total energy decreases steadily, or not. In Fig.~\ref{fig:spectraldensityseperation}(d), this is due to a change of the spectral density. Here we analyze this in more detail, with reference to the possibility of the system to behave as a heater, accelerator, or engine; see Fig.~\ref{fig:fig1totalsystem}(c).

Within each period, we define the efficiency of the thermal machine
\begin{align}
\label{eq:efficiency}
&\eta=\frac{\bar{J_e^L}-\bar{J_e^R}}{\bar{J_e^L}}=1-\frac{\bar{J_e^R}}{\bar{J_e^L}}
\end{align}
where $\bar{J_e^\nu}$, with $\nu=L,R$, is the average energy current from each bath over one period $\mathcal{T}=2\pi/\omega_{\rm dri}$, and $\bar{{J}^\nu_e} = (1/\mathcal{T})\int_{t_0}^{t_0+\mathcal{T}} J^\nu_e (\tau)d\tau$ \cite{footnoteaverageJe}. We point out here that both $\bar{J_e^L}$ and $\bar{J_e^R}$ used to compute $\eta$ below are positive (flowing from the left bath into the system, and from the system into the right bath) unless stated otherwise. Thus when $\eta>0$, the system is an engine, and it becomes an accelerator when $\eta<0$ [Fig.~\ref{fig:classification}(a)]. When the sign of $\bar{J_e^L}$ changes, then the system turns into a heater.

In Fig.~\ref{fig:classification}(a) we study the efficiency as a function of the left bath temperature $T_L$  for different coupling strengths $\gamma$. The hot bath needs to have a minimal temperature for the system to behave like an engine, and this temperature increases for larger $\gamma$. It is also apparent that for larger $\gamma$ the efficiency is reduced. \new{We interpret this as an effective broadening of the energy spectra of both the system and the baths, thus rendering the energy transfer less efficient.}
In Fig.~\ref{fig:classification}(b) we plot the efficiency versus the value of the peak resonance of the spectral density of the right bath for different coupling strengths. We observe that for frequencies around \new{$\omega^L_r\approx \omegas + \omega_{\rm dri}$, the system has a peak of efficiency as an engine, which is due to the matching of resonance}. \new{Consistently with Fig.~\ref{fig:classification}(a),} the efficiency is reduced by larger $\gamma$, and the region in which the system performs as an engine is reduced. In Fig.~\ref{fig:classification}(c) we plot more explicitly the efficiency versus coupling strength for different driving amplitude $\delta$. As expected, for larger $\gamma$, and for smaller $\delta$, the efficiency becomes smaller. In Fig.~\ref{fig:classification}(d) we depict the average current over a period versus the amplitude of the driving $\delta$. With darker colors we show larger left bath temperatures, while with continuous lines we depict the current with the left bath, and with the dashed line we depict the current with the right bath. Observing the lighter lines, we observe that for larger driving amplitude $\delta$ the current for the left bath may become negative, meaning that the system is turning into a heater. For intermediate left-bath temperature, both currents can be positive, but since the dashed line is above the continuous one, the system is an accelerator. For the darker lines, the highest $T_L$ considered, the system is an engine because both currents are positive and the current from the hot bath is larger than the one towards the cold bath.
Hence, given a large driving amplitude $\delta$, by varying the temperature of the left bath one can turn the system from a heater to an accelerator to an engine. \new{One thus requires a large enough temperature difference to obtain an engine, however a too large amplitude of the driving can result in a too strong non-linear response which can turn the system into an accelerator.}

\subsection{Signatures of bath correlations}\label{sec:correlations}

In Fig.~\ref{fig:classification} we observed that for larger system-bath coupling strength $\gamma$ one obtains lower efficiency. The thermofield transformation with chain mapping that we use to evolve our system allows us to study the overall system under unitary evolution. This implies that one can obtain information on the correlations building between the system and the baths, and even correlations between the baths. To do so, we start from the correlation matrix $\sigma_T=\langle c_i^\dagger c^{}_j \rangle$, where $c^{}_j$ represents either the operator $a_{\rm S}$ or $b_k^\nu$ in Eq.~(\ref{eq:totalHambeforeTC}). From this correlation matrix we can build the reduced correlation matrices $\sigma_{S+B_i}$ between the system and the bath $i$, $\sigma_{B_L+B_R}$ between the two baths, and $\sigma_{S}$ for the system and $\sigma_{B_i}$ for each bath.

The correlations between each one of the baths and the system, and between two baths, are defined by the relative entropies \cite{Vedral2002}
\begin{align}
\label{eq:correlationmatrices}
&C(\sigma_{S+B_i}||\sigma_{S}\otimes\sigma_{B_i}) \\ \nonumber &= {\rm Tr}\sigma_{S+B_i}\left[\ln\sigma_{S+B_i} - \ln\left(\sigma_{S}\otimes\sigma_{B_i}\right)\right] \,\,(i=L,R)  \\ \nonumber
&C(\sigma_{B_L+B_R}||\sigma_{B_L}\otimes\sigma_{B_R}) \\ \nonumber &= {\rm Tr}\sigma_{B_L+B_R}\left[\ln\sigma_{B_L+B_R} - \ln\left(\sigma_{B_L}\otimes\sigma_{B_R}\right)\right].
\end{align}
\begin{figure}[ht]
\centering
\includegraphics[width=1.0\columnwidth, draft=false]{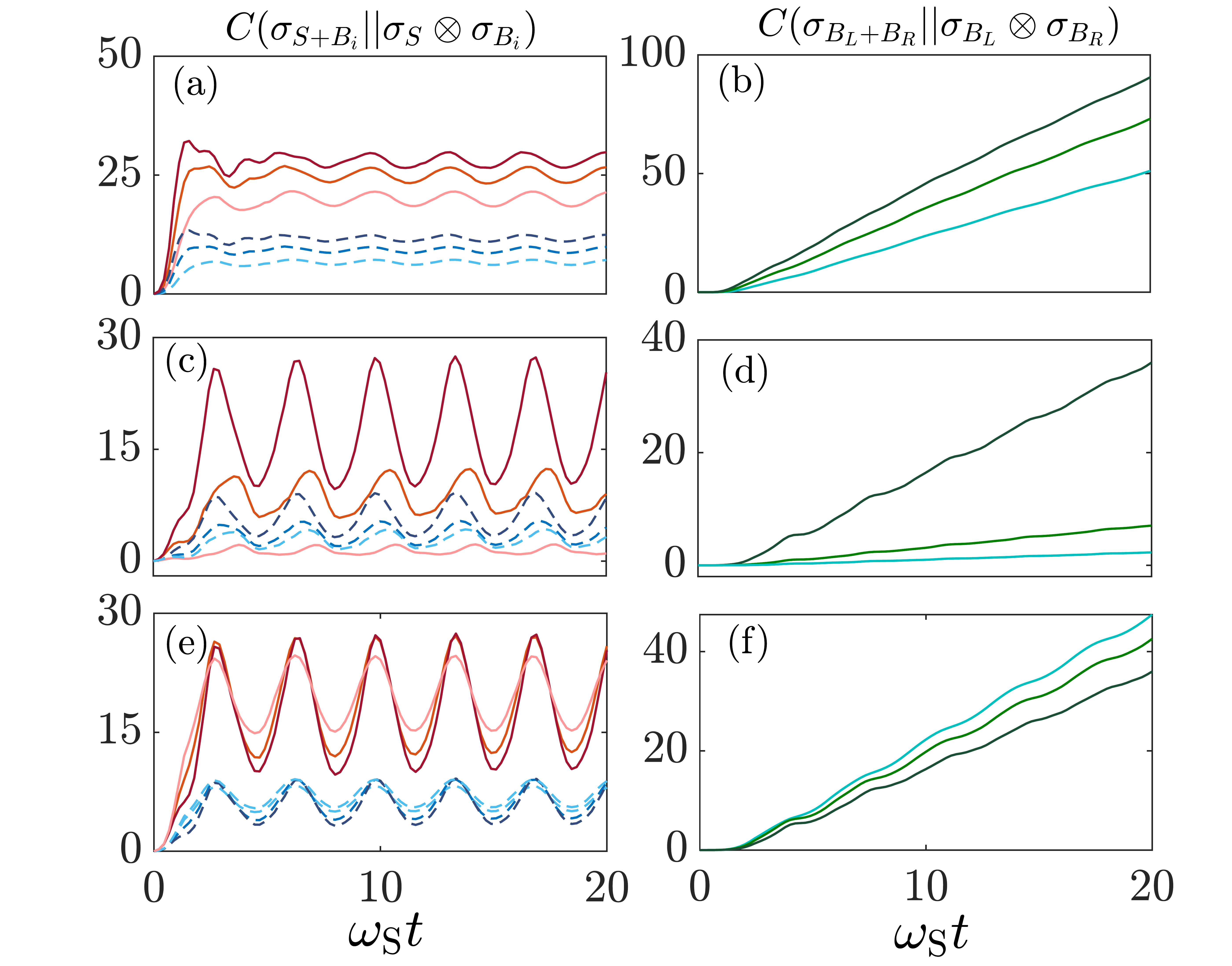}
\caption{\label{fig:correlationmatrices}Measure of system-bath and bath-bath correlations as a function of time: (a), (c), and (e) are correlations between the system and each bath \textcolor{black}{$C(\sigma_{S+B_i}||\sigma_{S}\otimes\sigma_{B_i})\,\,(i=L,R)$}, and (b), (d), and (f) are correlations between two baths $C(\sigma_{B_L+B_R}||\sigma_{B_L}\otimes\sigma_{B_R})$. In (a), (c), and (e) the solid lines are for $C(\sigma_{S+B_L}||\sigma_{S}\otimes\sigma_{B_L})$, and the dashed lines are for $C(\sigma_{S+B_R}||\sigma_{S}\otimes\sigma_{B_R})$. (a) and (b) Correlations as a function of system-bath coupling strength $\gamma = 0.25$, $0.5$, $1$ (darker color for larger value). Other parameters: $T_L=20T_0$, $\delta=0.1\omegas$. (c) and (d) Correlations as a function of the left bath temperature $T_L=3T_0$, $5T_0$, $10T_0$ (darker color for larger value). Other parameters: $\gamma=0.25$, $\delta=0.7\omegas$. (e) and (f) Correlations as a function of the driving amplitude $\delta=0.3\omegas$, $0.5\omegas$, $0.7\omegas$ (darker color for larger value). Other parameters: $T_L=10T_0$, $\gamma=0.25$. For all panels, we have used $T_R=T_0$, $\omega_{\rm dri}=0.5\omegas$, $\omega_c=2.75\omegas$, $A_L=A_R=0.0125$, and the measurement step is $0.2$.}
\end{figure}

In Fig.~\ref{fig:correlationmatrices} we look for the effect of system-bath coupling strength, the temperature bias, and the driving amplitude on the correlations. The left column represents the correlations between the system and the left bath $C(\sigma_{S+B_L}||\sigma_{S}\otimes\sigma_{B_L})$ (red solid curves), and between the system and the right bath $C(\sigma_{S+B_R}||\sigma_{S}\otimes\sigma_{B_R})$ (blue dashed curves), while the right column represents the correlations between both baths $C(\sigma_{B_L+B_R}||\sigma_{B_L}\otimes\sigma_{B_R})$ (green curves). In all panels, the system-bath correlations reaches periodic steady states after long enough evolution. On the other hand, the bath-bath correlations keep growing, on average linearly, in the long time. It is evident from Fig.~\ref{fig:correlationmatrices}(a) and (b) that a stronger system-bath coupling strength leads to larger system-bath correlations, and the bath-bath correlations increase much faster when the system-bath coupling is stronger. Also, it is worth mentioning that in panel (a) the periodic steady-state correlation between the system and the left bath is always larger than the one between the system and the right bath. This is consistent with the results of $\eta>0$ (more energy transported into the system from the left bath than those out of the system into the right bath) from Fig.~\ref{fig:classification}(c) when the system behaves as an engine under the same parameters chosen here. In Fig.~\ref{fig:correlationmatrices}(c) and (d), we consider the effect of the temperature bias, and we observe that larger temperature bias gives larger system-bath correlations and faster increase of the bath-bath correlation. We notice, however, that when the temperature bias is small such that the system is an accelerator [$T_L=3T_0$, lighter-colored curves in panel (c)], the periodic correlation between the left bath and the system is smaller than the one between the right bath and the system. This indicates that the magnitude of the current plays an important role in the magnitude of the correlations and their growth.
Finally, in Fig.~\ref{fig:correlationmatrices}(e) and (f), we consider the effect of the driving amplitude $\delta$. We find that the driving amplitude also gives rise to varying the amount of correlations that build up between the system and the bath, and how fast they can grow between baths.
However, the exact dependence of these effects on $\delta$ is also a function of the other parameters in the system and the bath.
Overall, from our numerical investigations, shown in Fig.~\ref{fig:correlationmatrices} or not, we have found that an increase in current tends to correspond to an increase of the rate at which the correlations between the baths grow.

\section{Conclusions}
\label{sec:conclusions}

In this work we have considered a periodically driven harmonic oscillator coupled to two baths of harmonic oscillators at different temperatures. We have used a numerical approach based on thermofield transformation with chain mapping \cite{Wilson1975,Plenio2010a,ChinPlenio2010,ChinPlenio2014,deVegaBanuls2015} which allows to simulate exactly the system and the baths unitarily up to a finite time. In our analysis we are able to study long enough times such that the system reaches a steady state and we characterize it.
We find that by tuning both the system and the bath parameters, one can turn the system from a heater to an accelerator and also to an engine. In our investigation we focused on the coupling strength, on the resonance frequency of the baths and the spread of the baths' density of states, and also on the driving amplitude and bath temperatures.
Since our approach goes beyond the weak-coupling approximations, we are able to also study correlations forming between the baths and the system, and also between the hot and cold baths.

In the future one could model the system as a larger, interacting setup, so as to study the interplay between interactions and the corresponding phases of matter, periodic driving and boundary driving from the baths. For this type of work, the thermofield transformation with chain mapping described herein could be readily used. \new{Fermionic systems with fermionic baths would also be of particular relevance to solid-state systems applications.} More studies should also be done to better understand the role and importance of the correlations that build between the baths, which one would expect naturally when considering the overall setup of system plus baths as a large unitary system, but which are neglected in typical weak-coupling master equation approaches.

\begin{acknowledgments}

\end{acknowledgments}
We acknowledge fruitful discussion with V. Balachandran, D. Gelbwaser-Klimovsky, Z. L. Lim, V. Scarani, B. Xing and X. Xu. D. P. acknowledges support from the Ministry of Education of Singapore AcRF MOE Tier-II (Project No.~MOE2018-T2-2-142). The computational work for this article was partially performed on resources of the National Supercomputing Centre, Singapore (NSCC) \cite{NSCC}.

\bibliographystyle{unsrt}

\begin{thebibliography}{100}
\bibitem{fn1} A fourth way in which the system can function is as a refrigerator, in which heat goes from the cold to the hot bath and work is done on the system \cite{BuffoniCampisi2019}.
\bibitem{fn2} Here, $\Delta \omega$ is set to be $0.0025\omegas$, and there is a maximum cut-off of the frequency for each bath at $\omega_c$.
\bibitem{fn_length_chain} In all calculations hereafter we use $N_{\text{chain}}=360$.
\bibitem{footnoteaverageJe} The average energy current from both baths is computed from the data of the last two periods and the last period, and again compare them and assure that the relative error from them is of the order $10^{-5}$.
\end{thebibliography}

\end{document}